\documentclass[aps,prl,twocolumn,superscriptaddress, longbibliography, floats]{revtex4-1}

		\usepackage{graphicx}
		\usepackage{amsmath}
        \usepackage[usenames,dvipsnames]{color}
        \usepackage{pdfcolmk}

\begin{document}

    \title{Adequacy of Si:P Chains as Fermi-Hubbard Simulators}

\author{Amintor Dusko}
\email{amintor.dusko@gmail.com}
\affiliation{Instituto de F\'{\i}sica, Universidade Federal do Rio de Janeiro, Caixa Postal 68528, 21941-972 Rio de Janeiro, Brazil}
\affiliation{Instituto de F\'{\i}sica, Universidade Federal Fluminense, Brazil (present address)}

\author{Alain Delgado}
\affiliation{Department of Physics, University of Ottawa, Ottawa, Ontario, Canada}

\author{Andr\'{e} Saraiva}
\affiliation{Instituto de F\'{\i}sica, Universidade Federal do Rio de Janeiro, Caixa Postal 68528, 21941-972 Rio de Janeiro, Brazil}

\author{Belita Koiller}
\affiliation{Instituto de F\'{\i}sica, Universidade Federal do Rio de Janeiro, Caixa Postal 68528, 21941-972 Rio de Janeiro, Brazil}

		\date{\today}

%		\maketitle
		
\begin{abstract}
\section{abstract}
The challenge of simulating many-body models with analogue physical systems requires both experimental precision and very low operational temperatures. Atomically precise placement of dopants in Si permits the construction of nanowires by design. We investigate the suitability of these interacting electron systems as simulators of a fermionic extended Hubbard model on demand. We describe the single particle wavefunctions as a Linear Combination of Dopant Orbitals (LCDO). The electronic states are calculated within configuration interaction (CI). Due to the peculiar oscillatory behavior of each basis orbital, properties of these chains are strongly affected by the interdonor distance $R_0$, in a non-monotonic way.  Ground state (T=0K) properties such as charge and spin correlations are shown to remain robust under temperatures up to  4K for specific values of $R_0$. The robustness of the model against disorder is also tested, allowing some fluctuation of the placement site around the target position. We suggest that finite donor chains in Si may serve as an analog simulator for strongly correlated model Hamiltonians. This simulator is, in many ways, complementary to those based on cold atoms in optical lattices---the trade-off between the tunability achievable in the latter and the survival of correlation at higher operation temperatures for the former suggests that both technologies are applicable for different regimes.

\end{abstract}

		\maketitle
\section{Introduction}
Strongly interacting fermions are the main ingredient for some phenomena in the forefront of physics, such as high-$T_c$ superconductivity and topological phase transitions \cite{Hasan2010, Qi2011, Jotzu2014}. In one dimension, correlations may be identified through collective electronic behavior such as charge density wave (CDW), bond-order wave (BOW) and spin density wave (SDW)]\cite{Esslinger2010, Boll2016}. The complexity in describing correlated particles led to the idea of simulating these systems with artificial architectures mimicking many-body models such as the Hubbard Hamiltonian. Control over this Hamiltonian parameters requires a level of experimental precision achieved only recently with cold atoms in optical lattices. This experimental platform presents low tunneling probability for atoms in neighboring optical lattice sites, which sets a low energy scale for quantum effects.
Optical lattice based experiments obtained many-body manifestations as spin and charge correlations for 1D \cite{Boll2016} and 2D \cite{Cheuk2016, Parsons2016} lattices. In the 1D case, related to the present study, the required temperature is in the nanokelvin range.
{ Other promising proposals as the quantum simulation using a semiconductor quantum dot array were presented recently \cite{Hensgens2017}, and will become a competitive architecture once long quantum dots arrays become feasible.  }

In the last few years, the expertise in positioning dopants nanostructures in Si has evolved~\cite{Shinada2005, Weber2012, Prati2012, Prati2016, Salfi2016}. {The precision necessary for quantum applications with regard to impurity placement has pushed the development of these techniques to the point that atomic scale certainty is now a reality~\cite{Schofield2003, Fuechsle2012, Zwanenburg2013}. }Precise atomic chains of these impurities are fabricated and, as demonstrated theoretically here, they may constitute convenient simulators for the extended Hubbard model. Multi-valley effects are ubiquitous in Si and valley interference impacts the tunnel coupling. We investigate--within a realistic model--how electronic properties of donor nanowires in Si can be controlled by design to emulate Hubbard systems, even allowing for some disorder effects.

\section{Model and Methods}

These atomistic wires may extend throughout several nanometers, and a full description of the Si atoms would not be feasible. Instead, we describe the wire electronic states as a Linear Combination of Donor Orbitals (LCDO)~\cite{Dusko2016}. Each basis orbital is an effective mass Kohn-Luttinger(KL) variational wavefunction for the ground state (A$_1$ symmetry)~\cite{Kohn1955},
\begin{equation}
\label{eq:wave_function}
    \Psi_{\rm KL} ({\bf r}) ={\frac{1}{\sqrt{6}}} \sum_{\mu=1}^6 F_\mu ({\bf r}- {\bf R_i}) \phi_\mu ({\bf r}- {\bf R_i}),
\end{equation}
where ${\bf R_i}$ is the position vector of the substitutional donor at site $i$. The index $\mu = 1 \ldots  6$ labels the degenerate minima of the Si conduction band at   ${\bf k}_\mu$ along the six equivalent $\langle 100 \rangle$ directions, i.e. $\pm x, \pm y, \pm z$, $\left| {\bf k}_\mu \right|=k_0=0.85\left( \frac{2 \pi}{a_{Si}}\right)$, and $a_{Si}$ is the conventional Si lattice parameter \cite{Madelung2012}. In this approach, the influence of the Si host is explicitly included in the orbitals, allowing the investigation of longer chains than the conventional fully atomistic approach. When directly compared to experiments, this multivalley central cell corrected KL wavefunction leads to the correct single impurity spectrum~\cite{Saraiva2015}, single impurity wavefunction~\cite{saraiva2016} and two impurities spectra, both in the ionized states~\cite{gonzalez2014} and neutral excited states~\cite{dehollain2014}.

We use isotropic hydrogen-like envelopes, $ F_{\mu} ({\bf r})=F ({ r}) ={\left({\pi a^{*3}}\right)^{-1/2}} \, \exp\left( -r/a^*\right)$. These isotropic envelopes include a species dependent Bohr radius  $a^*$ obtained by considering a screened potential, affected by the Si host and  the donor singular potential. Isotropic envelopes do not incorporate the effective mass anisotropy around the conduction band minima in Si, which is not relevant here. Its validity, including for the current system, is discussed in Ref.~[\onlinecite{Saraiva2015}]. Screening is treated by considering a functional form for the Coulomb potential that interpolates the expected asymptotic behaviors $V(r\rightarrow0)={\pm e^2}/{4 \pi \epsilon_{0} r}$ and $V(r\rightarrow \infty)={\pm e^2}/{4 \pi \epsilon_{Si} r}$, where the $+(-)$ signal stands for electron-electron (electron-proton) potential. This transition between bare and screened point charge potentials occurs at a phenomenologically determined screening length $r^{*}$, such that the full potential reads
\begin{equation}
  V(r)=\pm\frac{ e^2}{4 \pi r}\left[\frac{1}{\epsilon_{Si}} + \left( \frac{1}{\epsilon_{0}} - \frac{1}{\epsilon_{Si}} \right)e^{-{r}/{r^{*}}} \right],
\end{equation}
where $\epsilon_{Si}$ is the Si static dielectric constant, and $\epsilon_{0}$ the vacuum susceptibility. In what follows, we consider this screening both on single and two particle Hamiltonian terms. In the electron-electron Coulomb potential, we take $r^{*}=0.1$\,nm, a typical value for a Si environment~\cite{Saraiva2015}.

 \begin{figure*}[!]
		\includegraphics[clip,width=0.8\textwidth]{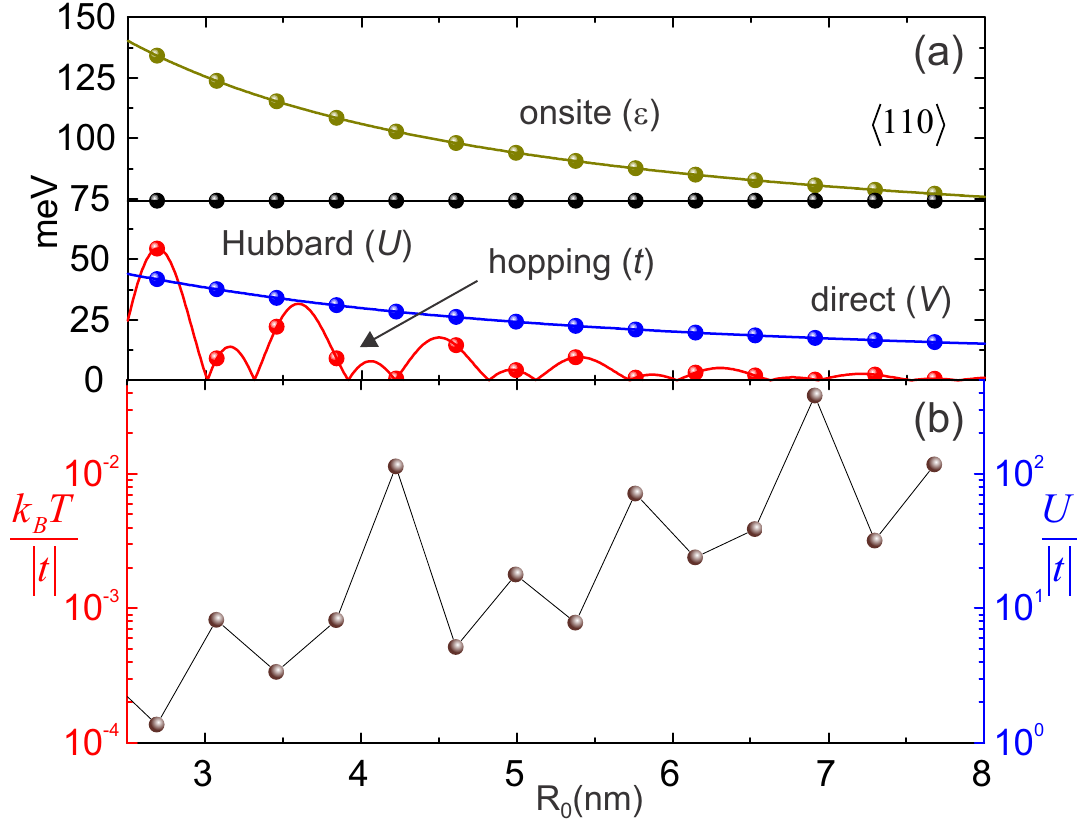}
		\caption{\label{fig4} (a) Absolute values of the calculated Hamiltonian parameters Hubbard ($U$), direct ($V$), multi-valley hopping ($t$) and onsite energy ($\varepsilon$). Spheres mark the values of $R_0$ allowed for donor pairs in Si along [110].
(b) Relative values of the Hubbard energy and $k_B T$ at temperature $T= 100$mK with respect to the hopping absolute value. The sharp oscillations are due to $|t|$ alone, as the others are constants. The temperature of $100$mK is attainable for experiments performed under dilution refrigerators. Small $k_B T/\left|t\right|$ and large $U/\left|t\right|$ favor the experimental implementation of the proposed simulator (see text).
}
\end{figure*}

We study a first nearest neighbors Hamiltonian written in the LCDO basis~\cite{Dusko2016}. Defining the creation and annihilation operators $c^+_{i,\sigma}$ and $c_{i,\sigma}$ for an electron at the orbital centered in ${\bf R_i}$ with a spin projection $\sigma$ along a quantization axis, and the corresponding number and charge density operators $n_{i,\sigma}=c^+_{i,\sigma} c_{i,\sigma}$ and $\varrho_{i}=n_{i,\uparrow} + n_{i,\downarrow}$, the Hamiltonian reads
\begin{equation}\label{eq:hamiltonian}
  H  =\sum_{i,\sigma}\varepsilon_{i} n_{i,\sigma}+\sum_{\langle i, j \rangle, \sigma}t_{ij} c^+_{i,\sigma} c_{j,\sigma}
                + \sum_{i}U_{i} n_{i,\uparrow}n_{i,\downarrow} + \sum_{\langle i, j \rangle}V_{ij} \varrho_{i} \varrho_{j}.
\end{equation}

This is readily recognizable as the extended Hubbard Hamiltonian \cite{Hubbard1963, gebbhard1997}, with parameters $\varepsilon_{i}$ (onsite), $t_{ij}$ (hopping), $U_{i}$ (Hubbard) and $V_{ij}$ (direct). Analytic expressions for theses parameters are given in the Supplemental Material~\cite{SM}, calculated values for a set of interdonor distances are presented in Fig.~{\ref{fig4}(a).

These parameters, which are consistent with typical orders of magnitude obtained experimentally~\cite{Shinada2005, Prati2012, Salfi2016}, support the idea that chains of dopants in Si constitute a strongly correlated system. Figure~\ref{fig4}(b) shows that the ratio between the onsite Coulomb repulsion and the tunnel coupling is $U/t\approx1$ to $100$, which ranges from the metallic regime, through the Mott insulator transition up to the strong localization driven by interactions. Still, the tunnel coupling $t$ is strong enough that quantum fluctuations are dominant over thermal excitations even at dilution fridge temperatures $T\approx 100$mK. At this temperature, we have $k_B T/t\approx10^{-4}$ to $10^{-2}$. Even at liquid He temperatures, this ratio is lower than 0.1 for all ranges of interdonor separation suggested here. In comparison, state-of-art cooling techniques applied to cold atoms still are not able to achieve ratios lower than $k_B T/t\approx0.2$.

There is strictly no long range order in a one dimensional chain. Still, a rich variety of low-temperature electronic ordering tendencies appears at the range of parameters discussed here. Regardless of the strongly non-monotonic behavior with $R_0$, as a general trend small distances favour a CDW phase, while increasing $R_0$ we pass through a BOW phase and a SDW phase is favored at larger distances (see the Supplemental Material~\cite{SM}). We investigate signatures of these many-body effects from charge and spin correlations.

\section{Results}
We focus on the Hilbert subspace of neutral (half-filling) chains with $N_S=8$ sites, periodic boundary conditions and zero total spin projection $S^z_{tot}=0$ [sketched in Fig.~{\ref{fig1}]. We arbitrarily set the quantization axis to $z$, without regard to its significance with relation to the crystallographic directions. In the absence of external magnetic fields, this choice is arbitrary and the solutions to this Hamiltonian is invariant under a rotation of spin states.  Donor positions are assigned at evenly spaced (${\bf R_0}$) substitutional atomic sites in Si along a [110] crystalline direction. The many-body state is described within the Configuration Interaction (CI) framework~\cite{Gallagher1997, Wensauer2004, Ozfidan2013, Ozfidan2015} and diagonalized exactly. Since Si is a material with very low spin-orbit coupling and no piezoelectric phonons, it is reasonable to assume that spin relaxation times are much longer than all other time scales involved in the experiment, so that thermalization does not remove the system from the $S_z=0$ subspace.

 \begin{figure*}[!]
		\includegraphics[clip,width=0.8\textwidth]{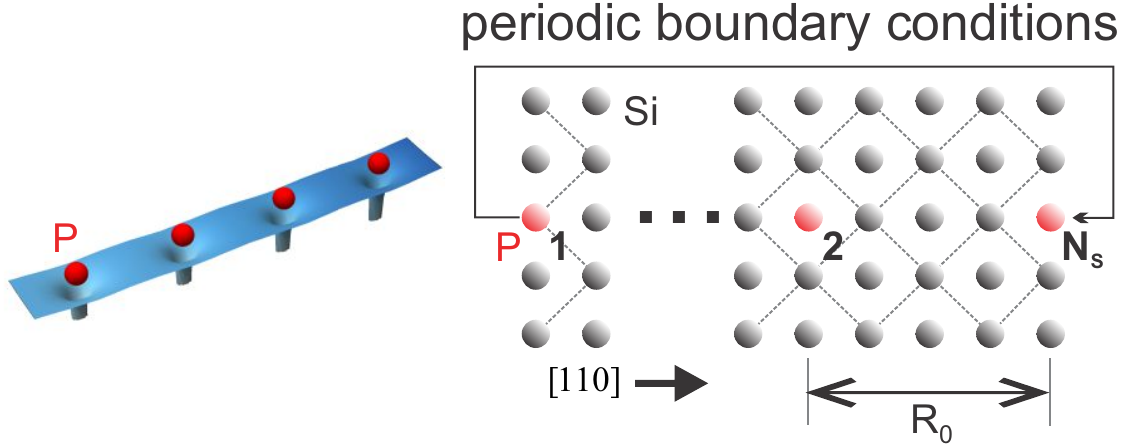}
		\caption{\label{fig1} Representation of a 1D chain of $N_S$  dopants in Si aligned along the $[110]$ direction and with interdonor separation $R_0$. We adopt periodic boundary conditions,  connecting site $j=N_S$ with site $1$.
}
\end{figure*}

Given a set of local operators $\{A_i\}$ acting on site $i$, a pair correlation function may be defined with $i=1$ taken as the reference site
\begin{equation}
  \mathcal{F}_{1,j}(A)=\sum_{n=1}^{N_\ell}w_n\left({\langle A_1 A_j \rangle_n - \langle A_1 \rangle_n \langle A_j \rangle_n }\right)
\end{equation}
where the average is taken over the thermally excited equilibrium occupations, $N_\ell$ is the total number of states, $w_n$ is the Boltzmann weight of a level $n$ at a given temperature $T$ and $\langle \cdots \rangle_n = \langle \Phi_n \left|  \cdots \right| \Phi_n \rangle$ is the expectation value of the operator for $|\Phi_n\rangle$, the $n^{\rm th}$ eigenstate of $H$.

We define the dimensionless correlation function $\mathcal{A}_j(A)=\mathcal{F}_{1,j}(A)/\mathcal{F}_{1,1}(A)$, so that for any $T$ the self-correlation is $\mathcal{A}_{1} = 1$, while   $\mathcal{A}_{j} = 0$ when the values of $A$ at sites $1$ and $j$ are completely uncorrelated.
It is now straightforward to define  charge-charge [$\mathcal{C}_{j}=\mathcal{A}_{j}(\varrho)$] and spin-spin [$\mathcal{S}_{j}=\mathcal{A}_{j} (S^z)$] electronic correlations with the total spin at site $j$ component along the quantization axis defined as $S_j^z = \frac{1}{2} \left( c^+_{j,\uparrow} c^{}_{j,\uparrow} -c_{j,\downarrow}^+ c^{}_{j,\downarrow} \right)$.

Mermin-Wagner theorem~\cite{Mermin1966} states that one dimensional chains at finite temperatures can not sustain a long range order that breaks continuous symmetries. Still, the range of the pair correlations is a valuable figure of merit for the appearance of collective behavior in low-dimensional systems. We initially discuss charge correlations as a function of temperature T and interdonor spacing $R_0$.

Results for the extreme cases T=0K and T=300K, are shown in the Supplemental Material~\cite{SM}.
At absolute zero, $\mathcal{C}_{j}$ is restricted to holes in the nearest neighbors of the reference site, namely $j=2$ and $j=8$, with strong localization due to the Mott mechanism. Any other pair $\{1,j\}$ remains essentially uncorrelated.
{ Results for 300K
are presented merely as an illustration of high temperatures limit, when even the nearest neighbors'
correlations are lost. Neither the proposed device nor the model developed here are suitable for this temperature range.  }
At room temperature, even the nearest neighbors correlations are lost.

 \begin{figure*}[!]
		\includegraphics[clip,width=0.8\textwidth]{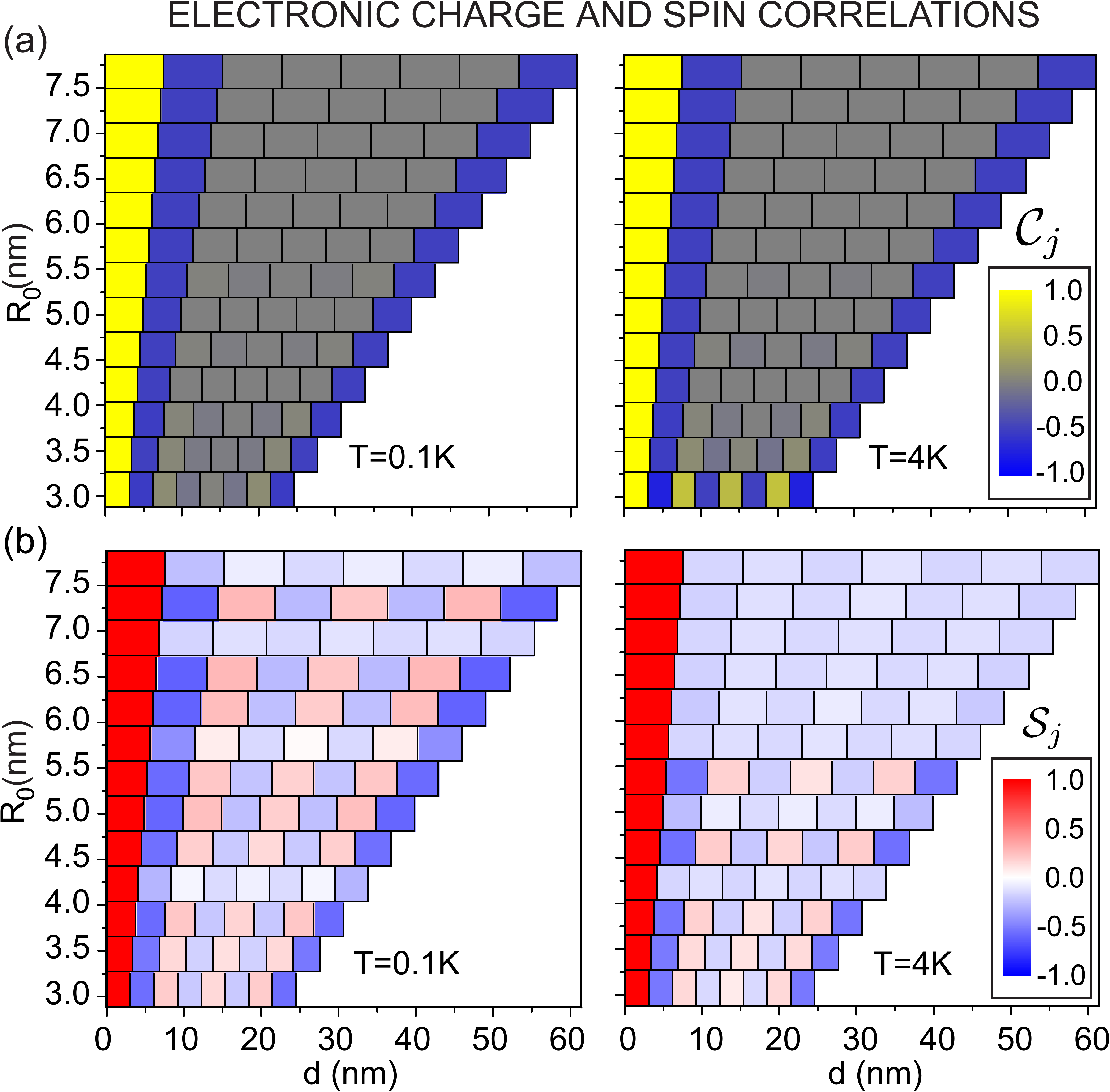}
		\caption{\label{fig2} Electron-electron correlations with respect to site $j=1$ (therefore $\mathcal{C}_1=\mathcal{S}_1 = 1$, see text). (a) Calculated charge-charge correlations for the indicated temperatures. Each box shows $\mathcal{C}_j$ (see color code) as a function of interdonor distance ($R_0$) and position along the chain ($d$). Note that for T=4K and $R_0\sim 3$\,nm the charge-charge correlations are temperature activated. (b) Same as (a) for calculated spin-spin correlations. An anti-ferromagnetic (AF) behavior up to fourth neighbors is clearly observed--the fifth, sixth and seventh neighbors pairs are equivalent to third, second and first pairs, respectively, due to periodic boundary condition. Note that the magnetic behavior is not monotonic with $R_0$.
}
\end{figure*}

Fig.~\ref{fig2} shows results at two experimentally achievable low temperatures. For T=0.1K the 0K results are essentially reproduced while at T=4K and $R_0\sim 3$\,nm there is an alternation in $\mathcal{C}_{j}$ among successive $j$'s, i.e., a temperature-activated delocalization. This indicates that, for this particular $R_0$,  states with metallic character within $k_B T\sim 0.3$ meV of the ground state dominate the Boltzmann average.

Spin correlations propagate further into the chain.
The antiferromagnetic correlations among nearest neighbors hint at the establishment of a SDW phase, as expected for this range of parameters in the $U$ vs $V$ phase diagram. For T=0K, the antiferromagnetic-like behavior is observed along all $j$ (see Supplemental Material~\cite{SM}), still with stronger correlations with neighboring sites $(j=2,8)$.
Results at room temperature, as for the charge, show no indication of correlation between sites. The T=0.1K and 4K results show a strong sensitivity of spin correlations with $R_0$--some specific distances sustain the ground state antiferromagnetic tendency while others show very weak correlation signatures. This is a consequence of the oscillatory behavior of the hopping $t$ with $R_0$ as can be observed in Fig.~\ref{fig4}(a). If the hopping is  small (large), correlations will be weaker (stronger). Thus, for T=0.1K and $R_0 = 6.53$\,nm  the chain is fully correlated, while for $R_0=6.91$\,nm the correlations vanish, reappearing for $R_0=7.30$\,nm at this temperature.

Unavoidable positional disorder impacts all Hamiltonian parameters. We estimate this effect in the electronic properties through a simple model for disorder where P donors can occupy any position inside a disk with radius $\delta=0.4$\,nm around a target substitutional site (see Fig.~\ref{fig3}(a)). With this uncertainty radius, each donor can occupy 5 positions. This level of uncertainty is realistically achieved for STM placement of donors \cite{Weber2012,Salfi2016}.

 \begin{figure*}[!]
		\includegraphics[clip,width=0.8\textwidth]{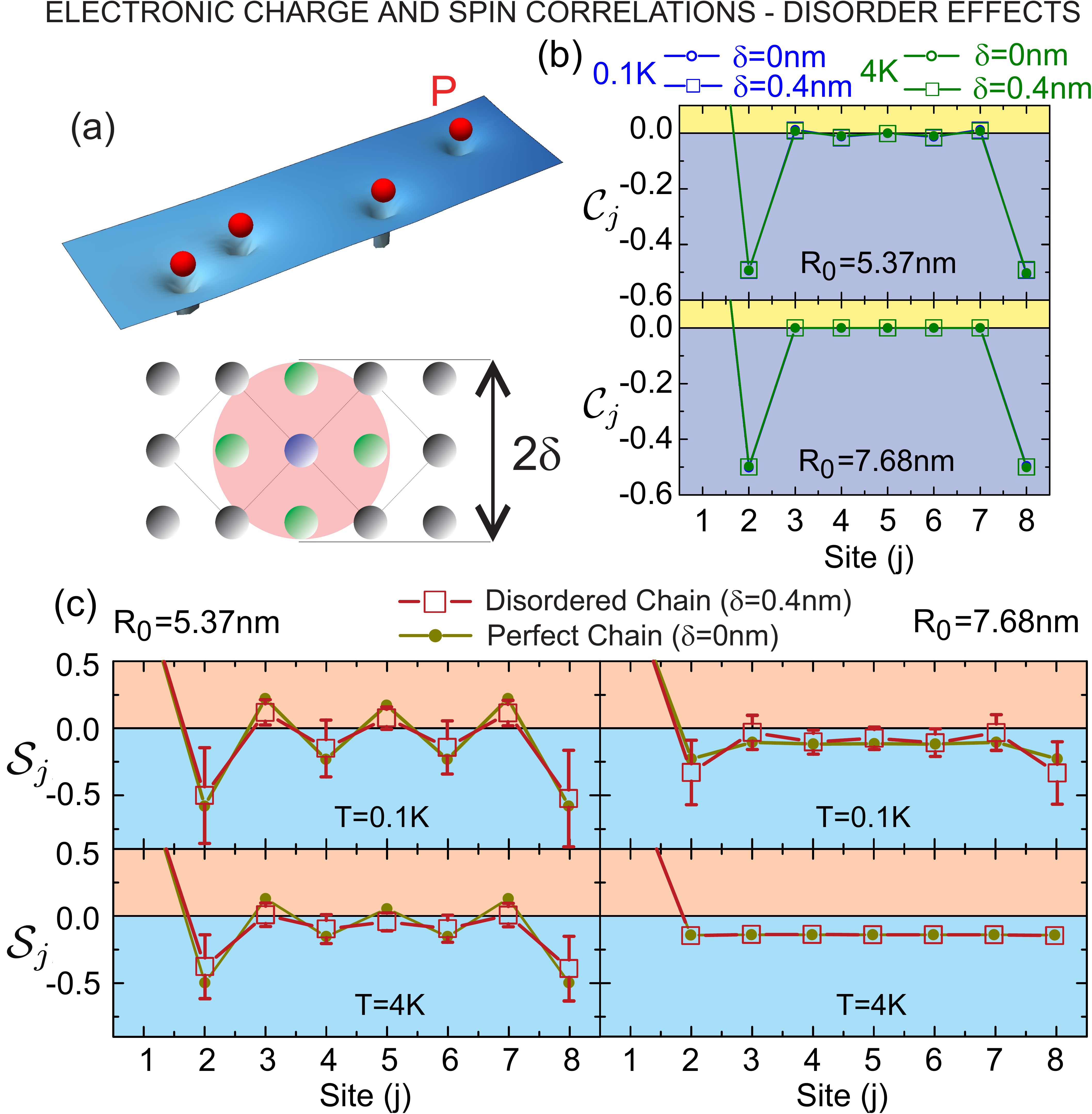}
		\caption{\label{fig3} (a) Model of 2D disorder: the $j^{\mathrm{th}}$ donor occupies the target (blue) or additional Si structure sites (green) within a distance $\delta$ from it. The figure corresponds to $\delta = 0.4$\,nm. (b) Disorder effects on  charge ($\mathcal{C}_j$) correlations at the indicated temperature, $\delta$, $R_0$ and site.
 Square data points and error bars give averages and standard error calculated over an ensemble of 1000 8-site disordered chain realizations.  (c) Same as (b) for $\mathcal{S}_j$ correlations.
 In all cases we compare disordered ($\delta = 0.4$\,nm) and ordered ($\delta = 0$\,nm -- circles) cases.  Note that $\mathcal{S}_j$ is more susceptible to disorder than $\mathcal{C}_j$.
}
\end{figure*}

These nanowires are quasi-1D chains where electrons can follow only one path \cite{Dusko2016}. To investigate effects of disorder and temperature we compare results for perfect and disordered chains for two nominal distances between target sites ($R_0$). { Both interdonor distances are chosen to be significantly larger than the range of disorder $R_0\gg \delta=0.4$ nm.} We choose the two distances sustaining ($R_0=5.37$\,nm) and loosing ($R_0=7.68$\,nm) the AF correlations at low $T>0$ [See Fig.~\ref{fig2}(b)].
Results for $\mathcal{C}_j$ shown in Fig.~\ref{fig3}(b) show that in all cases effects of the 2D disorder are mild, and do not affect significatively $\mathcal{C}_j$ general trends--even for temperatures up to 4K. Magnetic correlations, on the other hand, are more clearly affected by disorder, as seen in Fig.~\ref{fig3}(c). On average, disorder leads to less than $20\%$ reduction for non neglectable correlations.
The dispersion among spin correlations for individual realizations, indicated by the error bars in Fig.~\ref{fig3}, means that disordered samples may present sizeable correlations, eventually stronger than the ordered ones.
In real experiments, small chains are susceptible to this dispersion and may result in enhanced magnetic behavior.

\section{Discussions and Conclusions}

Perhaps the most important challenge for the implementation of the simulator described here is the measurement of the correlation function. While a direct measurement of charge and spin is possible~\cite{watson2015,buch2013} -- these are the basis of the Kane model of quantum computation -- it might be easier to extract these correlations from charge transport measurements~\cite{weber2014}.

The natural electronic correlations that appear in these chains may constitute an important resource for the study of many-body physics. It displays peculiar properties, constituting a unique example of a strongly interacting system with disordered tunnel coupling due to valley interference. This kind of random phase of the tunnel coupling element is the main ingredient in models displaying critical unitary statistics~\cite{Batsch1996,Cheng2016}.  Moreover, the ongoing development of nanofabrication capabilities suggest that on-demand models may be analogically simulated. For instance, the intricate phase diagram of the Fermi-Hubbard problem may be unveiled by spin- or density-resolved microscopy measurements\cite{Boll2016}.
Such application is under intensive investigation within cold atoms in optical lattices, and the present technology may complement these efforts. While not as easily tunable, the mass fabrication of circuits of donors adopting the know-how from available semiconductor technology would allow to chart the behavior of electrons over a wide range of attributes. The resilience of correlations in Si:P chains under relatively high temperatures suggests an attractive avenue for future experimental investigation.

{We have shown here that, up to currently accessible values of position disorder and temperature, dopant arrays in silicon preserve quantum correlations among atoms in a diluted chain. Our key point is that this system constitutes a robust implementation of the Fermi-Hubbard model in a semiconductor system with on-demand Hamiltonian parameters.}

\section{Acknowledgments}
This work is part of the Brazilian National Institute for Science and Technology on Quantum Information and made possible by the facilities of the Shared Hierarchical Academic Research Computing Network (SHARCNET:www.sharcnet.ca) and Compute/Calcul Canada.
The authors acknowledge CAPES (Science without Borders Program) for funding an 1-year stay for Amintor Dusko in Quantum Theory Group (uOttawa) and partial support from FAPERJ [E-26/202.915/2015 (BK)] and CNPq (fellowships).
Amintor Dusko thanks Marek Korkusinski and Isil Ozfidan for many fruitful discussions.
We would also like to thank Richard Scalettar, Thereza Lacerda Paiva, Rubem Mondaini and Peter Zoller for encouragement and discussions, and especially Pawel Hawrylak for following, discuss and support the development of this work.

\section{Competing Interests}
The authors declare that they have no competing financial interests.

\section{Contributions}
Numerical and analytical calculations were performed by A. Dusko., A. Delgado supervised the many-body methodology.  A. Dusko, A. Saraiva and B. Koiller discussed the work together and wrote the manuscript.
%
%\section{Funding}
%Partial funding from grant FAPERJ E-26/202.915/2015 (Belita Koiller)

\vfill
~

\bibliographystyle{naturemag}

\end{document}